\documentclass[aps,twocolumn,superscriptaddress]{revtex4-1}

\usepackage{amsmath}
\usepackage{mathtools} 
\usepackage{amssymb}
\usepackage{amsthm}
\usepackage{bm} 
\usepackage{upgreek} 
\usepackage{xcolor}
\usepackage{graphicx}
\usepackage{epstopdf}
\usepackage[colorlinks,linkcolor=blue,anchorcolor=blue,citecolor=blue,urlcolor=blue]{hyperref}

\newcommand{\nn}{{\nonumber}}
\newcommand{\bea}{\begin{eqnarray}}
\newcommand{\eea}{\end{eqnarray}}

\newcommand{\av}[1]{\left\langle #1 \right\rangle}

\newcommand{\md}{\mathrm{d}}
\newcommand{\me}{\mathrm{e}}
\newcommand{\w}{\omega}

\newcommand{\xvhs}{x_\textsc{\tiny VHS}}

\begin{document}

\title{Magnetotransport in overdoped La$_{2-x}$Sr$_x$CuO$_4$: a Fermi liquid approach}

\author{Rui-Ying Mao}
\affiliation{National Laboratory of Solid State Microstructures $\&$ School of Physics, Nanjing University, Nanjing 210093, China}

\author{Da Wang} \email{dawang@nju.edu.cn}
\affiliation{National Laboratory of Solid State Microstructures $\&$ School of Physics, Nanjing University, Nanjing 210093, China}
\affiliation{Collaborative Innovation Center of Advanced Microstructures, Nanjing University, Nanjing 210093, China}

\author{Congjun Wu} \email{wucj@physics.ucsd.edu}
\affiliation{Department of Physics, University of California, San Diego, California 92093, USA}

\author{Qiang-Hua Wang} \email{qhwang@nju.edu.cn}
\affiliation{National Laboratory of Solid State Microstructures $\&$ School of Physics, Nanjing University, Nanjing 210093, China}
\affiliation{Collaborative Innovation Center of Advanced Microstructures, Nanjing University, Nanjing 210093, China}

\begin{abstract}
Recently, several experiments on La$_{2-x}$Sr$_x$CuO$_4$ (LSCO) challenged the Fermi liquid picture for overdoped cuprates, and stimulated intensive debates \cite{Bozovic_review_2019}.
In this work, we study the magnetotransport phenomena in such
systems based on the Fermi liquid assumption.
The Hall coefficient $R_H$ and magnetoresistivity $\rho_{xx}$
are investigated near the van Hove singularity $\xvhs\approx0.2$
across which the Fermi surface topology changes from hole- to electron-like.
Our main findings are:
(1) $R_H$ depends on the magnetic field $B$ and drops from positive to negative values with increasing $B$ in the doping regime $\xvhs<x\lesssim0.3$;
(2) $\rho_{xx}$ grows up as $B^2$ at small $B$ and saturates at large $B$, while in the transition regime a ``nearly linear'' behavior shows up.
Our results can be further tested by future magnetotransport experiments in the overdoped LSCO.
\end{abstract}
\maketitle

After more than three decades of efforts, there still exist many mysteries
in cuprate superconductors, partly because the superconductivity
arises from two fundamentally different states: the parent undoped Mott
insulating state and the heavily overdoped metallic state \cite{Lee2006}.
Close to the Mott insulator side, 
only doped holes contribute to charge transport and the carrier density
equals the doping level $x$ (per Cu).
On the other hand, in the heavily overdoped metallic region,
the total carrier density is expected to change to $1+x$.
Such an anticipated transition from $x$ to $1+x$ was reported to occur at a critical doping level $x^*$ by measuring the normal state Hall number $n_H$
in YBa$_2$Cu$_3$O$_y$ (YBCO) \cite{Badoux2016} and La$_{1.6-x}$Nd$_{0.4}$Sr$_x$CuO$_4$ (Nd-LSCO) \cite{Collignon2017}
under strong magnetic field.
Here, $n_H$ is defined as $\frac{V}{eR_H}$ with $R_H$ the Hall coefficient and $V$ the volume per Cu, 
such that the sign of $n_H$ indicates the carrier type.
In combination with many other experiments \cite{Tallon2001, Keimer2015, Proust2019}, the sharp transition of $n_H$ from $x$ to $1+x$ is
possibly driven 
by an underlying quantum critical point (QCP) beneath the superconducting dome \cite{Eberlein2016, Storey2016, Maharaj2017, Verret2017, Mitscherling2018, Sachdev2018}.

However, as for the case of Nd-LSCO, there exists a puzzle: at $x>\xvhs\approx0.22$ where  $\xvhs$ is the doping when the Fermi energy reaches the van Hove singularity (VHS) \cite{Matt2015,Leyraud2017},  $n_H=1+x$ is in conflict with the prediction of the Lifshitz-Azbel-Kagonov theory \cite{Lifshitz1957, Abrikosov1988}.
It states that in the strong magnetic field limit, the Hall number $n_H$ should be given by the electron number enclosed by the Fermi surface (FS), i.e., $n_H=-(1-x)$ where the minus sign indicates the carriers are electrons rather than holes.
On the other hand, if the magnetic field value $B$ is not large enough,
the Hall number is not determined by the Luttinger volume \cite{Luttinger1960} directly but depends on the FS curvature \cite{Ong1991}, hence, $n_H=1+x$ is not anticipated either.

In fact, the transition from $x$ to $1+x$ has not been observed in La$_{2-x}$Sr$_x$CuO$_4$ (LSCO).
Similar to Nd-LSCO, there also exists a VHS at $\xvhs\approx0.18\sim0.20$
as observed by the angle-resolved photo emission spectroscopy (ARPES)
experiment \cite{Yoshida2006}.
One consequence of the VHS is that the normal state Hall coefficient $R_H$ decreases smoothly with doping and finally drops to negative values at $x\approx0.3$ in the weak magnetic field limit \cite{Hwang1994, Nishikawa1994, Tsukada2006}.
In fact, the smooth behavior of $R_H$ upon doping is consistent with
the measurements of the upper critical field \cite{Rourke2011}, superfluid density \cite{Bozovic2016} and resistivity \cite{Cooper2009} in LSCO.
Certainly, some experiments reported
possible QCP signatures such as the insulator-to-metal transition around the optimal doping $x_c\approx 0.16$ \cite{Boebinger1996}, or, the vanishing of the stripe/nematic order \cite{Cheong1991, Wu2017, Wen2019}, accompanied by a peak \cite{Balakirev2009} or upturn of $n_H$ \cite{Laliberte2016} upon doping at low temperatures under strong magnetic fields.
Another evidence of QCP comes from the observation of the linear magnetoresistivity at $0.16<x<0.19$ which was attributed phenomenologically to the linear scattering rate $\tau^{-1}\propto B$ \cite{Giraldo-Gallo2018} similar to the Planckian dissipation \cite{Zaanen2019}.
However, such an explanation is based on an  assumption that the $B$-dependence only comes from $\tau^{-1}$, which still needs more careful exploration.
Nevertheless, these experiments, together with the observation of nematicity \cite{Wu2017} and the very low superfluid density \cite{Bozovic2016} has stimulated a debate whether the metallic states in
overdoped cuprates are Fermi liquids or not. \cite{Bozovic_review_2019}

Motivated by these experimental progress, we employ the
Chambers' semiclassical theory \cite{Chambers1952,Chambers1990} to
study the Hall coefficient and magnetoresistivity for the general
values of the magnetic field $B$.
This study is based on the semiclassical cyclotron orbits of
quasiparticles on the Fermi surface.
It should be pointed out that the frequently used result of $R_H=\frac{1}{qn}$
is incorrect
unless in the strong field limit \cite{Lifshitz1957,Abrikosov1988},
otherwise it should be determined by the FS curvature in the weak field
regime as pointed out by Ong \cite{Ong1991}.
Based on the Chambers' formula, our calculations show the Hall
coefficient changes sign as increasing the field strengths in the range
of doping $\xvhs<x\lesssim0.3$.
Furthermore, there exists a ``nearly linear'' magnetoresistivity
at intermediate field strengths, especially when doping is
close to the VHS.
These results are consistent with the known experiments, indicating
that the magnetotransport properties of the overdoped LSCO could
still be described by the Fermi liquid theory.

\begin{figure}
\includegraphics[width=0.45\textwidth]{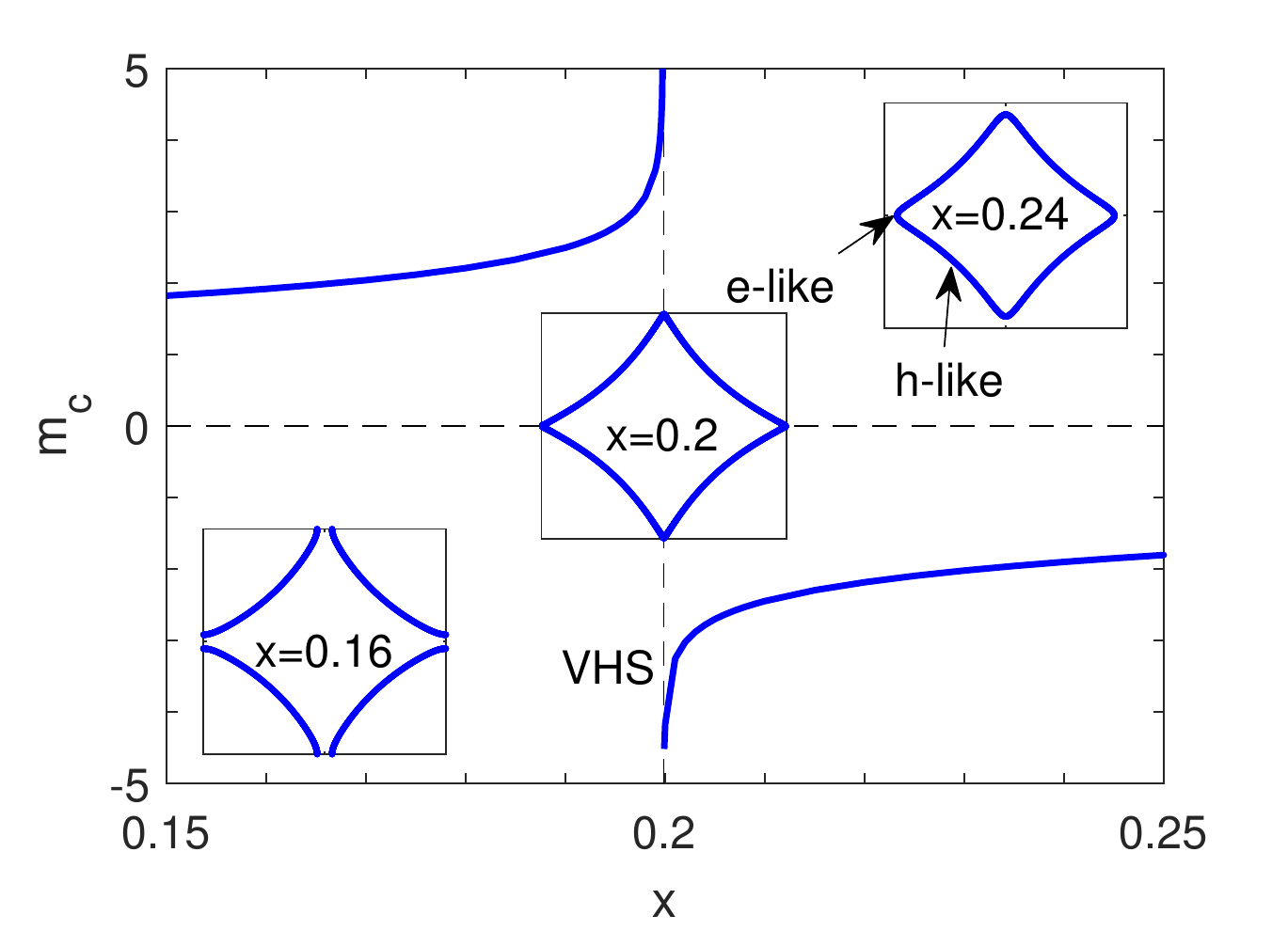}
\caption{Cyclotron mass $m_c$ (in unit of $\frac{\hbar^2}{a^2t}$) v.s. $x$.
The VHS is shown as the vertical dashed line.
The three insets show the FS below ($x=0.16$), almost at ($x=0.20$),
and above ($x=0.24$) the VHS filling.
Two arrows indicate hole- and electron-like FS segments in
the case of $x=0.24$, respectively.
}
\label{fig:FS}
\end{figure}

To capture the band structure of the overdoped LSCO, we adopt the
single-band model on the square lattice,
\begin{align}
H=-\sum_{\av{ij}\sigma} \left(t_{ij}c_{i\sigma}^\dag c_{j\sigma}+h.c. \right) -\mu \sum_{i\sigma}c_{i\sigma}^\dag c_{i\sigma} ,
\end{align}
where $t_{ij}$ is the hopping between sites $i$ and $j$, and
$\mu$ is the chemical potential.
To be specific, we denote $t$, $t'=-0.12t$ and $t''=0.06t$ as the first, second and third nearest neighbour hoppings obtained by fitting
the ARPES experiments \cite{Horio2018}.
When $\mu$ matches the VHS,
$\mu_\textsc{\tiny VHS}=4t'-4t''$ corresponding to
the doping $\xvhs\approx0.197$.
Three typical Fermi surfaces near the VHS are plotted in Fig.~\ref{fig:FS}.
At $x<\xvhs$, the FS is hole-like surrounding $(\pi, \pi)$, i.e.,
the corner of the Brillouin zone (BZ).
At $x>\xvhs$, the FS changes to surround the center of the BZ,
which is globally electron-like.
Nevertheless, the local curvatures of the FS segments change signs
from electron-like in the antinodal region (close to $(\pi,0)$ and $(0,\pi)$)
to hole-like in the nodal region (close to $(\pm\frac{\pi}{2},\pm\frac{\pi}{2})$),
leading to a multi-component feature \cite{Ong1991}.
Which picture (global or local) is more relevant for the Hall experiments
is an interesting question.

The key lies in the ratio of the scattering lifetime $\tau$ relative to the cyclotron period $\+T$.
The cyclotron frequency $\w_c=\frac{2\pi}{\+T}$ is determined by the cyclotron mass
$m_c=\frac{1}{2\pi}\frac{\partial\+S}{\partial\varepsilon}$ ($\+S$ is the area surrounded by the cyclotron orbit with energy $\varepsilon$)
through the relation $\w_c=\frac{eB}{|m_c|}$.
If $\tau$ is much larger than $\+T$, i.e., $\w_c\tau\gg1$, the cyclotron
motion completes the entire orbit leading to the global picture
of the electron-like FS.
Consequently, the Hall number $n_H$ is anticipated to be the electron number surrounded by the FS \cite{Lifshitz1957} according to the Luttinger theorem \cite{Luttinger1960}, i.e., $n_H=1+x$ for $x<\xvhs$ and $n_H=-(1-x)$ for $x>\xvhs$, respectively.
Meanwhile, the resistivity should be roughly proportional to the cyclotron mass $|m_c|$.
In contrast, if $\w_c\tau\ll1$, the quasiparticles have no chance to ``see'' the whole FS without scattering and thus the local picture is preferred.
As a result, we cannot identify $n_H$ as the carrier density directly
and the resistivity should be determined by the band mass $m^{-1}_{\alpha\beta}=\frac{\partial^2\varepsilon}{\partial k_\alpha \partial k_\beta}$ rather than the cyclotron mass $|m_c|$.

In order to obtain a unified description connecting the above two limits,
we adopt the semiclassical Chambers' formula
\cite{Chambers1952, Chambers1990, Abrikosov1988},
\begin{align}\label{eq:chambers}
\sigma_{\alpha\beta}=\frac{e^3B}{(2\pi)^2}&\int \md\varepsilon \left(-\frac{\partial f_0}{\partial \varepsilon}\right) \nn\\
&\int_0^\+T\md t v_\alpha(t) \int_{-\infty}^t\md t' v_\beta(t') \me^{-(t-t')/\tau} ,
\end{align}
where $v_\alpha=\frac{\partial \varepsilon_k}{\partial k_\alpha}$ depends on $t$ through the relation $\0k(t)$.
Eq.~\ref{eq:chambers} is derived based on the Boltzmann's transport equation
by considering the cyclotron motion perpendicular to $\0B$.
In cuprates, such a semiclassical picture can be justified by the
observations of quantum oscillations \cite{Sebastian2015} and cyclotron resonances \cite{Post2020}, which mean quasiparticles remain
coherent in magnetotransports.
We focus on the situation with $\0B$ perpendicular to the $ab$-plane
as in many experiment setups.

At low temperatures, only one cyclotron orbit, i.e., the FS, is needed to be considered.
The electron motion is determined by the Lorentz force $\hbar\dot{\0k}=-e\0v_k\times\0B$. 
After the velocity $\0v(t)$ is obtained (in this work by numerics), which satisfies the periodic condition $\0v(t)=\0v(t+\+T)$,  it can be derived that \cite{Maharaj2017}
\begin{eqnarray}
\sigma_{\alpha\beta}&=&\frac{e^3B}{(2\pi)^2} \frac{1}{1-\me^{-\+T/\tau}}\nn\\
&\times& \int_0^{\+T}\md t v_\alpha(t)
\int_{t-\+T}^t\md t' v_\beta(t') \me^{-(t-t')/\tau}.
\label{eq:sigma}
\end{eqnarray}
Then the Hall coefficient $R_H$ and magnetoresistivity $\rho_{xx}$ follow directly,
\begin{align}
R_H=\frac{\sigma_{xy}}{(\sigma_{xx}^2+\sigma_{xy}^2)B},\quad \rho_{xx}=\frac{\sigma_{xx}}{(\sigma_{xx}^2+\sigma_{xy}^2)} .
\end{align}
Based on Eq. \ref{eq:sigma}, $\frac{\sigma_{\alpha\beta}}{\tau}$ are
functions of $\w_c\tau$, yielding the Kohler's relations \cite{Ziman1960, Chambers1990}: $R_H=F(\w_c\tau)$ and $\rho_{xx}\tau=G(\w_c\tau)$
where $F$ and $G$ are functions of $\w_c\tau$.

Both the weak and strong field limits have been extensively studied in literature. When $\w_c\tau\ll1$, due to the exponential factor $\me^{-(t-t')/\tau}$, the $t'$-integral mainly comes from $t'\approx t$. Therefore, we can expand $v_\beta(t')=v_\beta(t)+\frac{\partial v_\beta}{\partial t}(t'-t)$ and substitute it into Eq.~\ref{eq:chambers}, giving
\begin{align} \label{eq:weakB}
J_\alpha&=e^2\tau \int_{\0k}\left(-\frac{\partial f_0}{\partial \varepsilon_\0k}\right)v_\alpha(\0k)v_\beta(\0k)E_\beta \nn\\
&+e^3\tau^2\int_{\0k}\left(-\frac{\partial f_0}{\partial \varepsilon_\0k}\right)v_\alpha(\0k)\frac{\partial v_\beta(\0k)}{\partial k_\gamma}E_\beta \left(\0v_\0k\times\0B\right)_\gamma ,
\end{align}
which is the same as the result obtained by Kubo formula \cite{Voruganti1992} and widely used in previous works \cite{Eberlein2016, Storey2016, Verret2017, Mitscherling2018}.
On the other hand, in the strong field limit $\w_c\tau\gg1$, the exponential factor $\me^{-(t-t')/\tau}$ can be approximated by unity, leading to $\sigma_{xy}=\frac{qn}{B}$ where $q=e$($-e$) when $m_c>0$($<0$) and $n$ counts the electron number surrounded by the FS \cite{Lifshitz1957, Abrikosov1988}. Then, $R_H=\frac{1}{qn}$ follows immediately due to the scaling behavior of $\sigma_{xx}\propto (\w_c\tau)^{-2}$.
Unfortunately, no simple results exist for $\rho_{xx}$ in the
strong field limit in general \cite{Ziman1960}.

\begin{figure}
\includegraphics[width=0.52\textwidth]{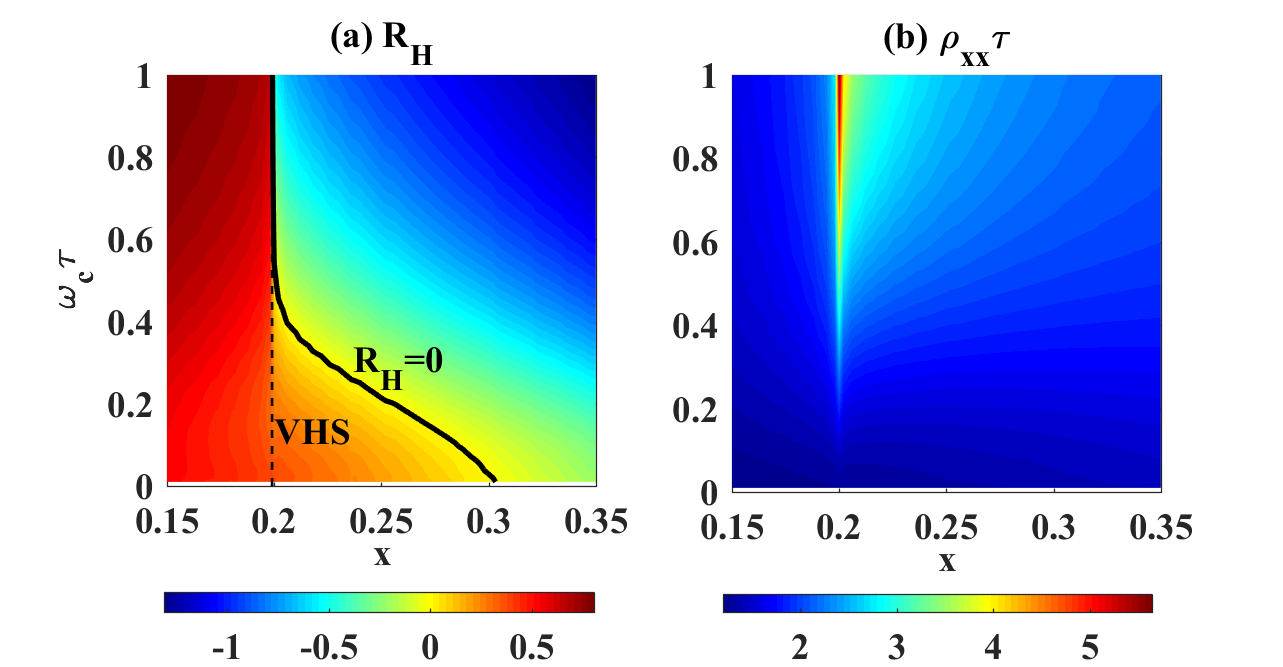}
\caption{Phase diagrams of $R_H$ (a) and $\rho_{xx}\tau$ (b) as functions of $x$ and $\w_c\tau$. The color encodes the values of $R_H$ (in unit of $\frac{a^2c}{e}$) and $\rho_{xx}\tau$ (in unit of $\frac{\hbar^2c}{e^2t}$). In (a), the contour of $R_H=0$ is shown as the solid line which merges to the VHS (dashed line) at large $\w_c\tau$.}
\label{fig:phasediagram}
\end{figure}

Our main results are shown in Fig.~\ref{fig:phasediagram}.
At first glance, both $R_H$ and $\rho_{xx}\tau$ show significant
dependence on $\w_c\tau$, exhibiting different behaviors from
free electrons in the Drude theory.
In the strong field limit $\w_c\tau\rightarrow\infty$,
$R_H$ changes exactly at $x=\xvhs$ reflecting the topological change of the FS and $\rho_{xx}\tau$ diverges similar to the behavior of an open orbit.
On the other hand, in the weak field limit $\w_c\tau\rightarrow0$, both $R_H$ and $\rho_{xx}\tau$ evolve smoothly with $x$.
The mismatch between the above two limits leads to many interesting
phenomena.
In the following, let us make detailed discussions about $R_H$ and $\rho_{xx}$, respectively.

Before moving forward, let us combine the scattering lifetime $\tau$
and the magnetic field $B$ into a dimensionless quantity $\lambda=\frac{eB\tau}{m^*}$ where $m^*=\frac{\hbar^2}{ta^2}$ is the band mass at the band bottom with
$a$ the lattice constant.
$\lambda$ is insensitive to doping and it equals $\omega_c \tau$
at the band bottom.
But close to the VHS, since $|m_c|$ is greatly enhanced, $\lambda\gg \omega_c\tau$.
The values of $\lambda$ are estimated below.
We choose $t\approx0.25$eV as obtained from ARPES \cite{Yoshida2006} and lattice constants $a\approx3.8\text{\AA}$, $c\approx6.6\text{\AA}$.
The scattering rate $\tau^{-1}$, however, is somewhat more difficult to determine. Upon doping, the interaction induced scattering may become weaker but the extrinsic disorder effect may become stronger.
$\tau^{-1}\sim5$meV is roughly estimated from the optical conductivity measurements in overdoped LSCO\cite{Mahmood2019},
in agreement with the very recent measurement in the optimally-doped
LSCO \cite{Post2020}.
With these parameters, $\lambda\approx0.01B$[Tesla].
Therefore, the magnetic field used in the experiments $B\leq80$T \cite{Cooper2009, Giraldo-Gallo2018} corresponds to $\lambda \leq0.8$.

\begin{figure}
\includegraphics[width=0.52\textwidth]{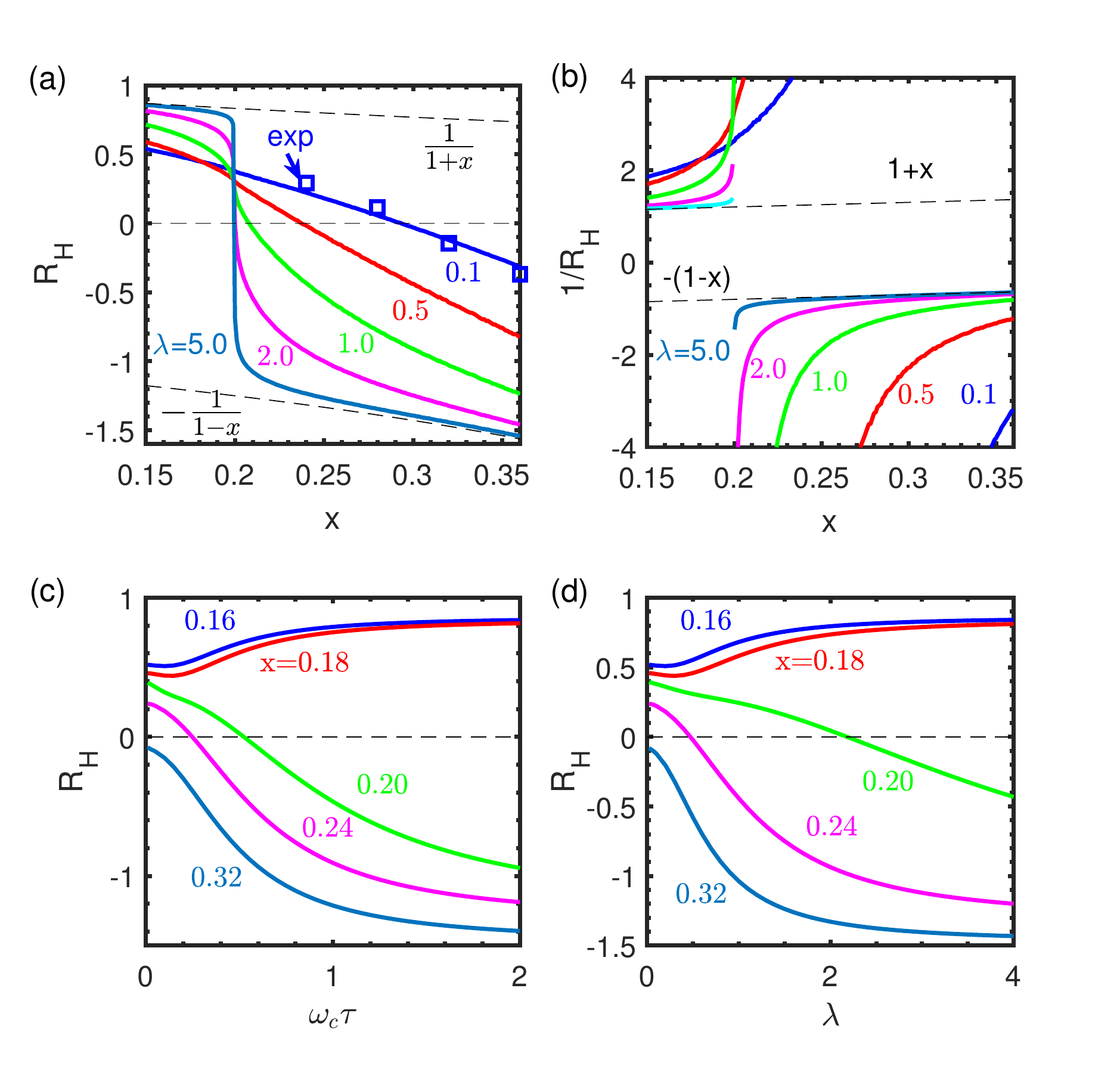}
\caption{
In ($a$) and ($b$), $R_H$ and $\frac{1}{R_H}$ are plotted v.s. the doping $x$, respectively.
Each curve corresponds to a different value of $\lambda=\frac{eB\tau}{m^*}$.
The experimental data of $R_H$ at $300$K \cite{Tsukada2006} in the weak field limit are shown as squares. The strong field limits $R_H=\pm\frac{1}{1\pm x}$ are also plotted as the dashed lines for comparison.
In ($c$) and ($d$), we plot $R_H$ v.s. $\w_c\tau$ and $\lambda$,
respectively, at fixed doping levels of $x$.
}
\label{fig:RH}
\end{figure}

Typical behaviors of $R_H$ are explicitly shown in Fig.~\ref{fig:RH}.
The most obvious feature is that its field dependence is very
different from that of free electrons with parabolic dispersions.
Fig.~\ref{fig:RH} ($a$) and ($b$) show that
only in the strong field limit $\lambda \gg 1$,
$R_H$ is given by counting the carrier numbers, {\it i.e.}, $eR_H=\frac{1}{1+x}$
and $-\frac{1}{1-x}$ when $x<\xvhs$ and $x>\xvhs$, respectively,
although the discontinuity of $R_H$ is smoothed by the finite lifetime.
However, this relation breaks down for weaker fields.
The experimental data of $R_H$ in LSCO \cite{Hwang1994, Nishikawa1994, Tsukada2006} are also plotted:
$R_H$ exhibits a significantly deviation from the scaling of
$-\frac{1}{1-x}$ and changes sign at $x\approx0.3$.
As shown in Fig.~\ref{fig:RH} ($c$), $R_H$ changes
significantly as varying $\omega_c\tau$.
The interesting regime lies in $\xvhs<x\lesssim0.3$, where $R_H$ drops from positive to
negative values as increasing the field strength, and finally
it saturates to the strong field limit $-\frac{1}{1-x}$.
Near the VHS, although $R_H$ changes sign at a finite value of $\omega_c\tau$, it requires a large field strength due to the
divergence of $m_c$, which may beyond the experimental
availability.
The field dependence of $R_H$ is replotted in  Fig.~\ref{fig:RH} ($d$)
in terms of $\lambda$.



\begin{figure}
\includegraphics[width=0.52\textwidth]{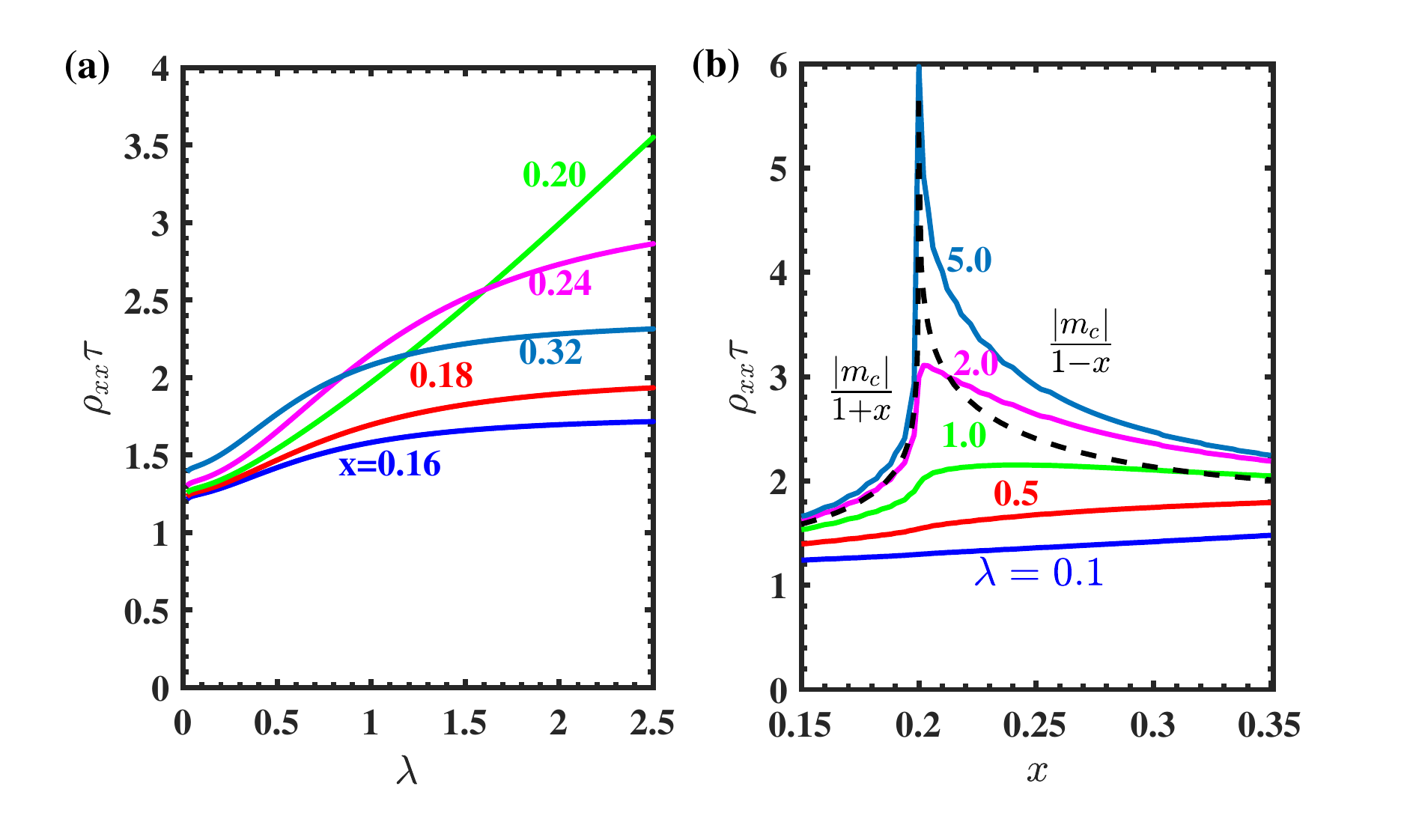}
\caption{Field (a) and doping (b) dependence of $\rho_{xx}\tau$. In (b), we also plot a rough estimation of $\rho_{xx}\tau\approx \frac{|m_c|}{1\pm x}$,
which qualitatively describes the behavior in the strong field limit.}
\label{fig:rho}
\end{figure}

Next we present the behavior of magnetoresistivity.
The relation of $\rho_{xx}$ v.s. $\lambda$ is shown in Fig.~\ref{fig:rho}($a$).
Away from the VHS, its behavior is standard as in usual metals \cite{Chambers1990}:
$\rho_{xx}\tau$ increases with $\lambda^2$ at $\lambda \ll 1$, 
saturates at $\lambda\gg 1$ with
$\lambda^{-2}$, and grows approximately linearly in between.
The $B^2$-behavior, {\it i.e.}, the $\lambda^2$-dependence, in the weak
field has been observed in LSCO \cite{Cooper2009, Giraldo-Gallo2018}.
The deviation from the $B^2$-behavior was found at $B\gtrsim30$T
corresponding to $\lambda \gtrsim0.3$ in experiments
\cite{Giraldo-Gallo2018}.
However, when the system is close to the VHS, due to the divergence of $m_c$, the
regime of linear growth is significantly enlarged, and the
magnetoresistivity is greatly enhanced.
The closeness to the VHS may provide an alternative explanation of the
linear magnetoresistivity observed in the experiments at $0.16<x<0.19$ \cite{Giraldo-Gallo2018}, rather than by the more exotic picture
of ``Planckian  dissipation", i.e., $\tau^{-1}\propto B$ \cite{Zaanen2019}.
Within our scenario, we would expect the tendency of
saturation of $\rho_{xx}$ around $B\sim100$T (corresponding to $\lambda
\approx 1$), which could be tested in the
future.
Fig.~\ref{fig:rho}(b) shows the doping dependence of $\rho_{xx}\tau$
at different values of $\lambda$.
At $\lambda\ll 1$, $\rho_{xx}\tau$ smoothly depends
on $x$ since the average of the band mass plays the dominant
role here.
As $\lambda$ increases, the cyclotron motion becomes more coherent.
The enhancement of $|m_c|$ drives the divergence of $\rho_{xx}\tau$
as $x$ approaching the VHS, which roughly follows
$\frac{|m_c|}{1\pm x}$ for $x<\xvhs$
and $x>\xvhs$, respectively.

Since the cyclotron resonance has been observed in the optimally doped
LSCO \cite{Post2020}, we argue that the role of cyclotron motion
cannot be neglected in the study of the magnetotransport behavior.
Certainly, in the strange metal region near the optimal doping,
our explanation
based on the Fermi liquid picture may not directly apply.
Nevertheless,
our results could be further tested in the more
heavily overdoped region and in higher fields.

In summary, we have performed a semiclassical study of the Hall coefficient and magnetoresistivity near the VHS in overdoped LSCO based on the Fermi liquid assumption.
Both $R_H$ and $\rho_{xx}\tau$ strongly depend on the magnetic field: as $B$ increases, $R_H$ changes sign from positive to negative values at $\xvhs<x\lesssim0.3$,
and $\rho_{xx}\tau$ grows up ``nearly linearly'' in the intermediate regime especially near the VHS.
Parts of the results are in good agreement with the known experiments and can be further checked whether the overdoped LSCO can be described by the Fermi liquid picture or not.


Before closing this paper, we provide some other remarks.
First, the band structure of LSCO is actually three-dimensional like
as we take into account of the out-of-plane hopping $t_z$, whose
value is at the same order of $t''$ as shown by the ARPES
measurement \cite{Horio2018}.
Although the effect of $t_z$ smears out the VHS, the field dependence
and sign change of $R_H$ are not expected to change qualitatively.


Second, VHS is not unique in LSCO but also in other hole-doped cuprates such as Nd-LSCO \cite{Matt2015}, Bi$_2$Sr$_2$CaCu$_2$O$_{8+\delta}$ (Bi2212) \cite{Kaminski2006}, Bi$_2$(Sr,La)$_2$CuO$_{8+\delta}$ (Bi2201) \cite{Kondo2004, Ding2019} and YBCO \cite{Gofron1994, Hossain2008}. Our study here indicates that the VHS needs more careful treatment when explaining their magnetotransport phenomena. Interestingly, Nd-LSCO and Bi2201 have the similar FS with LSCO and thus similar $B$- and $x$-dependence of $R_H$ should be observed experimentally.


At last, although $R_H$ obtained in the Fermi liquid picture is in good agreement with the experiments at high temperatures, its low temperature upturn is difficult to understand even in the highly overdoped region \cite{Hwang1994,Nishikawa1994, Tsukada2006}.
Moreover, taking the overdoped sample $x=0.23$ at $50$K as an example, the optical conductivity gives $\tau^{-1}\sim5$meV \cite{Mahmood2019}, giving rise to the theoretical value of $\rho_{xx}\sim5\mu\Omega\cdot\text{cm}$ at $B=0$, much smaller than the experimental value $\sim50\mu\Omega\cdot\text{cm}$ \cite{Cooper2009}.
In fact, this situation is similar to what happens in the superfluid density at zero temperature \cite{Bozovic2016}.
{\it The Fermi liquid picture severally underestimates $\rho_{xx}$ and $R_H$ at low temperatures and overestimate the superfluid density.}
This dilemma may be resolved by additionally taking the vertex corrections (not just band renormalization) into account, which is left as a following topic.


{\it Note added}. After completing the manuscript, we became aware of an interesting experiment on Nd-LSCO \cite{Grissonnanche2020} which measured both uniform and angle-dependent scattering rates, and also explained the linear magnetoresistivity by electron cyclotron motions.

We thank  Yao-Min Dai, Su-Di Chen and Ya-Yu Wang for helpful discussions.
This work is supported by National Natural Science Foundation of China (under Grant Nos. 11874205 and 11574134) and National Key Research and Development Program of China (under Grant No. 2016YFA0300401).


\bibliography{vhs_chambers}
\end{document}